\documentclass[12pt]{article}
\usepackage{epsfig,amsfonts,amssymb}
\usepackage{hyperref}
\usepackage{cite}
\input epsf.sty
\def\ZZZ{{\hbox{ Z\kern-1.6mm Z}}}
\def\RRR{{\hbox{ R\kern-2.4mm R}}}
\def\CCC{{\hbox{ C\kern-2.0mm C}}}
\def\zzz{{\hbox{z\kern-1mm z}}}

\newcommand{\qeq}{{\hbox{=\kern-2.3mm ? \kern.5mm }}}
\renewcommand{\qeq}{=}

\newcommand{\BB}{{\cal B}}

\newcommand{\AAA}{{\cal A}}

\newcommand{\OO}{{\cal O}}

\newcommand{\LL}{{\cal L}}

\newcommand{\wt}{\widetilde}

\newcommand{\NN}{{\cal N}}

\newcommand{\SSS}{{\cal S}}

\newcommand{\be}{\begin{equation}}
\newcommand{\ee}{\end{equation}}
\newcommand{\ben}{\begin{eqnarray}\displaystyle}
\newcommand{\een}{\end{eqnarray}}

\newcommand{\bea}[1]{\begin{eqnarray}\label{#1} }
\newcommand{\eea}{\end{eqnarray}}

\newcommand{\refb}[1]{(\ref{#1})}
\newcommand{\p}{\partial}

\def\one{{\hbox{ 1\kern-.8mm l}}}
\def\zero{{\hbox{ 0\kern-1.5mm 0}}}

\usepackage{cite}

\begin{document}

\baselineskip 24pt

\begin{center}
{\Large \bf Coulomb Branch of the Lorentzian 
Three Algebra Theory
}

\end{center}

\vskip .6cm
\medskip

\vspace*{4.0ex}

\baselineskip=18pt

\centerline{\large \rm   Sergio Cecotti$^a$ and 
Ashoke Sen$^b$ }

\vspace*{4.0ex}

\centerline{\large \it $~^a$Scuola Internazionale Superiore
di Studi Avanzati}

\centerline{\large \it via Beirut 2-4
I-34100 Trieste, ITALY}

\vspace*{4.0ex}

\centerline{\large \it $~^b$Harish-Chandra Research Institute}

\centerline{\large \it  Chhatnag Road, Jhusi,
Allahabad 211019, INDIA}

\vspace*{1.0ex}
\centerline{E-mail:  cecotti@sissa.it, 
sen@mri.ernet.in, ashokesen1999@gmail.com}

\vspace*{5.0ex}

\centerline{\bf Abstract} \bigskip

We analyze the coulomb branch of the non-unitary
Lorentzian three algebra
theory that has been proposed as 
a possible candidate for describing the
world volume theory of multiple M2-branes. In order that it
describes the theory of multiple M2-branes in flat eleven
dimensional space-time, the ghost fields must decouple and
the physical theory must be independent of the eight
coordinates of the moduli space representing the center of
mass coordinates of the branes. We show that the structure of
the Coulomb branch is consistent with this requirement.
While the full moduli space has the structure of a Lorentzian space
modded out by a Lorentz transformation, the physical subspace
has the correct structure of the moduli space of multiple 
M2-branes.
We also suggest a systematic procedure for testing the
consistency of the theory by
computing the higher derivative 
corrections to the effective action obtained by
integrating out the massive modes propagating in the loop.

\vfill \eject

\baselineskip=18pt

Since the discovery of a general class of $(2+1)$ dimensional
superconformal field theories due to Bagger, Lambert and
Gustavson\cite{0611108,0709.1260,0711.0955,0712.3738,0802.3456}, 
following earlier work
of  \cite{0411077,0412310}, there has been 
much activity in 
constructing and analyzing different types of three algebra
theories and other types of superconformal field theories
in three dimensions\cite{0803.3218}-\cite{0806.1639}. 
While much has been learned, 
identifying the correct theory of multiple M2-branes
based on three algebra theories remains an open problem.

A recent proposal for the world-volume theory of multiple
M2-branes makes use of a Lorentzian three 
algebra\cite{0805.1012,0805.1087,0805.1202}, and as
a result the scalar field kinetic term in this theory is not
positive definite.\footnote{Subsequently 
refs.\cite{0806.0054,0806.0738}
proposed a modification of this theory that involves
gauging a certain symmetry of the original theory and freezing
the unwanted degrees of freedom. Our analysis deals with the
original proposal of \cite{0805.1012,0805.1087,0805.1202}.}
In particular
the proposed world-volume theory for $N$ M2-branes consists
of two sets of eight scalar fields $X^I_+$ and $X^I_-$ with
$1\le I\le 8$, and $SU(N)$ algebra valued fields $X^I$, 
$\AAA_\mu$
and $\BB_\mu$. Besides these there are fermion fields which
we shall ignore for simplicity; our analysis can be easily
generalized to include the fermions. The bosonic part of the 
action has the form:
\be \label{ee1}
\SSS = \int d^3 x\, 
\p_\mu X^I_+ \p^\mu X^I_- + \SSS_1(X^I_+, X^I, \AAA_\mu,
\BB_\mu)\, .
\ee
{}From this we see that the $X^I_\pm$ equations of motion
take the form
\be \label{ee2}
\p^2 X^I_+ = 0, \qquad \p^2 X^I_- = 
{\delta\SSS_1\over \delta X^I_+}
\, .
\ee
The indefinite kinetic term comes from the sector
containing the fields $X^I_\pm$, -- this has eight scalar
fields with wrong sign kinetic term which we shall refer
to as the ghost fields. Had they been free fields
one could simply drop an appropriate linear combination of
$X^I_\pm$ with negative kinetic term from the 
theory.\footnote{At the level of the equations 
of motion it is perfectly
consistent to drop the $X^I_-$ field since it does not appear
on the right hand side of the equation of motion or the various
symmetry transformation laws of any other
field.  Indeed there is a reduced three algebra that acts only
on the $X^I_+$, $X^I$, $\AAA_\mu$
and $\BB_\mu$ fields together with their fermionic
superpartners without any need to introduce the $X^I_-$
fields and its superpartner\cite{0804.1784}. The only problem is that
one cannot write down an action without the $X^I_-$ fields
that reproduces these equations
of motion.}
However eq.\refb{ee2} shows that while
$X^I_+$ obeys free field 
equations of motion, $X^I_-$ has an interaction term. The
interaction originates in
the $\SSS_1$ component of the action describing the 
coupling of the $X^I_+$ field to the rest of the fields. 
If this theory is to describe a theory of multiple M2-branes,
there must be a consistent procedure allowing us to put an
appropriate restriction that removes the eight ghost fields and
gives us a unitary theory. The other eight scalars arising out
of $X^I_\pm$ could then describe the center of mass motion
of the branes; for this it is necessary that the dynamics in the
restricted subspace is independent of the vacuum expectation value
labelling the center of mass coordinate. 
At present we do not know of any such
consistent procedure, -- the proposal of 
\cite{0806.0054,0806.0738} essentially removes both sets of
degrees of freedom and leaves no room for constructing the
center of mass degrees of freedom out of the $X^I_\pm$ fields.
We shall proceed by assuming that there exists
some consistent mechanism
to truncate the theory by removing the eight ghost fields (and their
superpartners) without destroying the superconformal invariance
of the theory.

Since the theory has no coupling constant, it seems to be
impossible to test this assumption using any approximation
scheme. In fact it is
not clear if such a theory can be subject to any test at all,
since in the absence of a coupling constant
any test will involve solving the theory exactly. 
There is however one possible test one can subject this theory
to: namely that on the Coulomb branch, which is expected
to describe separated branes, 
the theory must correctly
reproduce the flat moduli space dynamics of the M2-branes
after removal of the ghost fields.
In particular the 
independence of the theory on the center of mass degree of
freedom should be manifest.
This is what we shall try to verify.\footnote{Some 
aspects of the moduli space of these
theories have been studied before (see {\it e.g.}
\cite{0805.1087}), but to our knowledge
a detailed study, taking into account all the global identifications,
has not been performed. A general discussion on the
moduli space of Lorentzian three algebra theories from an
abstract algebraic viewpoint can be 
found in \cite{0805.4363}, but the relationship between the
moduli space considered there and the one being analyzed here
is not completely transparent.}
At the end we shall also
suggest the possibility of systematically extending this
analysis to include higher derivative corrections to the
effective action of the moduli fields
by staying in a domain of the moduli space
where the massive modes are sufficiently 
heavy.

Let us for simplicity consider the case of Lorentzian 3-algebra
based on $SU(2)$ gauge group, although the analysis can be easily
generalized to the case of $SU(N)$ groups. We shall consider the
Coulomb branch where the $X^I_\pm$ and
$X^I$ fields take non-zero
expectation values. On this branch the off-diagonal components
of various $SU(2)$ triplet fields become massive, and hence can be
ignored while studying the low energy effective action.
The massless fields arise from the diagonal components
of the $SU(2)$ triplet fields:
\be \label{e1}
X^I = \wt X^I \, \sigma_3, \quad \BB_\mu =
b_\mu \sigma_3, \quad \AAA_\mu = a_\mu \sigma_3
\, .
\ee
The fields $\wt X^I$, $b_\mu$ and $a_\mu$, together
with the $X^I_\pm$ represent the set of massless bosonic
fields.
Our goal is
to analyze the low energy theory involving these massless 
fields. Our analysis will follow closely that of 
\cite{0804.1114,0804.1256} for the
$SO(4)$ Bagger-Lambert theory. 
The bosonic part of the Lagrangian involving these fields,
in the convention of \cite{0805.1087}, is
given by
\be \label{e1.5}
\LL_0 = - (\p_\mu \wt X^I - X^I_+ b_\mu)^2 + \p_\mu X^I_+ 
(\p_\mu X^I_- - 2b_\mu  \wt X^I)
+ \epsilon^{\mu\nu\lambda} b_\lambda \, f_{\mu\nu}, 
\ee
where
\be \label{e1.6}
f_{\mu\nu} =\p_\mu a_\nu - \p_\nu a_\mu\, .
\ee
We can dualize the gauge fields $a_\mu$ by treating
$f_{\mu\nu}$ in \refb{e1.5} as free fields and adding
the term 
\be \label{e1.7}
\Delta \LL=
- {1\over 4\pi}\,
\epsilon^{\mu\nu\lambda} \p_\lambda\, \phi \, f_{\mu\nu}\, ,
\ee
to \refb{e1.5}. 
Since in the convention of \cite{0805.1087} that we are using,
the conventionally normalized $SU(2)$ gauge field is 
$2\AAA_\mu$,
$\phi$ is an angular variable with period
$2\pi$. If we use the equations of motion for
$\phi$, we get the Bianchi identity $\epsilon^{\mu\nu\rho}
\p_\mu f_{\nu\rho} = 0$. On the other hand we can use the
equation of motion for $f_{\mu\nu}$ first to get
\be \label{e1.8}
b_\mu = {1\over 4\pi}\, \p_\mu\phi\, .
\ee
Substituting this
in $\LL_0+\Delta\LL$ we get the new Lagrangian density
\be \label{e2}
\LL = - \left(\p_\mu \wt X^I - {1\over 4\pi}\, X^I_+ 
\p_\mu\phi\right)^2 + \p_\mu X^I_+ 
\left(\p_\mu X^I_- - {1\over 2\pi}\, \wt X^I\, \p_\mu\phi \right)\, .
\ee

The original Lagrangian was
invariant under the $\BB_\mu$ gauge transformation:
\ben \label{e3} 
X^I_+ &\to& X^I_+\, , \nonumber \\
X^I_- &\to& X^I_-+ Tr(M X^I) + {1\over 2} \, Tr(M^2) X^I_+\, ,
\nonumber \\
X^I &\to& X^I+M X^I _+\, , \nonumber \\
\BB_\mu &\to& \BB_\mu + \p_\mu M+ 2 \, i\, [\AAA_\mu, M]\, .
\een
The gauge transformations which preserve the form \refb{e1}
correspond to the choice
\be \label{e4}
M = \wt M \, \sigma_3\, .
\ee
Under this we have
\ben \label{e5}
X^I_+ &\to& X^I_+\, , \nonumber \\
X^I_- &\to& X^I_-+ 2\, \wt M \wt X^I +  \wt M^2 X^I_+\, ,
\nonumber \\
\wt X^I &\to& \wt X^I+\wt M   X^I _+\, , \nonumber \\
\phi &\to& \phi+ 4\pi \, \wt M\, .
\een
$\LL$ given in \refb{e2} is invariant under this gauge
transformation. Furthermore the Weyl group 
of the $SU(2)$ gauge transformations
associated with 
the gauge fields $\AAA_\mu$ also preserves
the form \refb{e1} and acts as a symmetry of $\LL$ via
the transformation:
\be \label{eweyl}
\wt X^I\to -\wt X^I, \qquad \phi\to -\phi\, .
\ee
Two field configurations related by the transformation
\refb{eweyl} must be identified.

Now consider a point in the moduli space where 
one or more components of $X^I_+$ acquire  large vacuum
expectation value. Using $SO(8)$ invariance we can choose
$\langle X^I_+\rangle=v\, \delta_{I 8}$. In that case we can 
use \refb{e5} to set the gauge $\wt X^8=0$. Assuming that
$|\langle \wt X^I\rangle| << v$
and taking the fluctuations of the normalized scalar fields
to be small compared to $v$ we can express the Lagrangian
density as
\be \label{elead}
-\sum_{I=1}^7 \p_\mu \wt X^I\p^\mu \wt X^I 
- {v^2\over 16\pi^2}\,
 \p_\mu \phi
\p^\mu\phi + \p_\mu X^I_+ \p^\mu X^I_- + \OO(v^{-1})\, ,
\ee
where $\OO(v^{-1})$ terms contain cubic and quartic
interaction terms in $\wt X^I$, $X^I_\pm$ and $v\phi$.
Eq.\refb{elead} shows that the $\phi$ direction 
represents a circle of
radius $\propto v=\langle X^8_+\rangle$. 
In particular for fixed $X^I_+$ and $X^I_-$, the coordinates
$\wt X^I$ and 
$\phi$ span the space $\RRR^7\times S^1$
with the $S^1$ having length $\propto v$, modded out
by the $\ZZZ_2$ exchange symmetry given in \refb{eweyl}. The
$v$ dependence of the $S^1$ radius 
is a remnant of the fact that $v$ determines the coupling constant
of the $SU(2)$ 
Yang-Mills theory that arises for coincident M2-branes
\i.e. for $Y^I=0$\cite{0805.1012,0805.1087,0805.1202}.
Indeed we could have  first integrated 
out the field
$b_\mu$ and then $\phi$, with the result of getting a kinetic term for the gauge field $a_\mu$ proportional to
 $\frac{1}{v^2}f_{\mu\nu}f^{\mu\nu}$\cite{0803.3218}. 
 This
apparently suggests
that the physical theory depends on $v$. It was however
suggested in \cite{0805.3930} that 
this apparent dependence of the theory on the vacuum 
expectation value of $X^8_+$ may be due to a wrong choice
of field variables.
In particular since $X^8_+$ is
part of the moduli of the theory, 
we could, for small fluctuations in $\phi$ and $Y^I$, 
interprete 
$X^8_+$ (or more generally
$\sqrt{X^I_+X^I_+}$) as the radial
position of the center of mass of the brane in a cylindrical
polar coordinate system, $2\phi$ as the
relative separation of the branes along the azimuthal 
angle,
and $\wt X^I$ for $1\le I\le 7$ as the relative separation of the
branes along the radial and cartesian directions.
In that case the apparent $v$ dependence of the size of the
$\phi$ circle can be attributed to the peculiarity of the
polar coordinate system.

Now \refb{elead} is not the full Lagrangian. If we expand the
full Lagrangian \refb{e2} in powers of $1/v$, then there will
be correction to \refb{elead} in the form of interaction terms,
suppressed by powers of $1/v$, and will apparently correct the
moduli space metric. If on the other hand the
interpretation suggested above -- that the Lagrangian actually
describes the dynamics of M2-branes in flat space together with
a decoupled ghost term -- is correct then the full Lagrangian
\refb{e2} must describe free fields. 
Furthermore, in the physical subspace, the apparent dependence of the
$\langle X^I_+\rangle$=constant slice on $\langle X^I_+\rangle$
must disappear with appropriate choice of coordinates.
We shall now show that this
is indeed true.
We can proceed in two ways, -- either choose $\phi=0$ 
gauge in \refb{e2} or work with gauge invariant fields.
Both lead to the same result; so let 
us follow the second approach. We define
\ben \label{e6}
Y^I_+ &=& X^I_+ \, , \nonumber \\
Y^I &=& \wt X^I - {1\over 4\pi}\,
X^I_+ \phi\, , \nonumber \\
Y^I_- &=&  X^I_- - {1\over 2\pi}\,
\phi \wt X^I + {1\over 16\pi^2}\, \phi^2 X^I_+\, .
\een
It is easy to see that these variables remain invariant under
the gauge transformation \refb{e5}. In terms of these
variables the Lagrangian density
\refb{e2} takes the form:
\be \label{e7}
\LL=-\p_\mu Y^I \p^\mu Y^I + \p_\mu Y^I_- \p^\mu Y^I_+\, .
\ee
Thus the Lagrangian describes a set of free fields. In order to
identify the degrees of freedom of the M2-branes we make 
 a further field redefinition
\be \label{e8}
Z^I_- = Y^I_- + a\, Y^I_+\, ,
\ee
for some constant $a$
and treat $Z^I_-$, $Y^I_+$ and $Y^I$ as free fields. Then we have
\be \label{e9}
\LL = -\p_\mu Y^I \p^\mu Y^I - a\, \p_\mu Y^I_+
\p^\mu Y^I_+ + \p_\mu Z^I_- \p^\mu Y^I_+\, .
\ee
For appropriate choice of the constant $a$, $Y^I_+$ and $Y^I$ can
be interpreted as the center of mass and relative coordinates
of the two M2-branes. $Z^I_-$ on the other hand is the unwanted
ghost field which needs to be dropped from the action.
In particular we could declare the physical moduli space 
to be the
$Z^I_-$=constant slice; a  rigid translation
symmetry  of $X^I_-$
that the complete theory 
possesses\cite{0805.1012,0805.1087,0805.1202} 
guarantees that the
final theory is independent of the choice of this constant
value of $Z^I_-$.\footnote{Note that if we had taken the
$Y^I_-$=constant slice instead of the $Z^I_-$=constant
slice, it would make no difference to the equations of
motion or the structure of the physical
moduli space discussed below, but will fail to give the
correct contribution to the energy momentum tensor from
the $Y^I_+$ field. Since $Y^I_+$ is a free field, we could
rectify this by simply adding to the energy momentum tensor
of the other fields the free field contribution from the
$Y^I_+$ field. This would be an entirely equivalent 
prescription.}

Let us now study the various identifications in the
$(Y^I_+, Y^I, Z^I_-)$ space.
Eq.\refb{eweyl} leads to the 
identification $Y^I\to - Y^I$. This reflects the effect of
symmetrization under the exchange of two M2-branes.
On the other hand the $\phi\to\phi+2\pi n$ translation gives
the identification
\be \label{e10.1}
Y^I_+ \to Y^I_+\, , \qquad
Y^I \to Y^I - {n\over 2}\, Y^I_+\, , \qquad
Y^I_- \to Y^I_- -   n\, 
Y^I + {n^2\over 4} Y^I_+\, ,
\ee
on the $(Y^I_+, Y^I, Y^I_-)$ variables and
\be \label{e10}
Y^I_+ \to Y^I_+\, , \qquad
Y^I \to Y^I - {n\over 2}\, Y^I_+\, , \qquad
Z^I_- \to Z^I_- -   n\, 
Y^I + {n^2\over 4} Y^I_+\, ,
\ee
on the $(Y^I_+, Y^I, Z^I_-)$ variables.
{}From the point of view of
the Lorentzian free theory in eqn.\refb{e9}, 
this can be seen, for fixed $I$, as an identification
under a Lorentz transformation
mixing the $Y^I$ with the $Y^I_\pm$ directions. 
This is suggestive of some mysterious relation to 
Matrix theory \cite{BFSS}, and it will be interesting to
explore any possible connection in more details.\footnote{Some
related analysis in $SO(4)$ Bagger-Lambert theory can be found
in \cite{0804.2186}.}
%
%\textbf{The form of the Lorentz boost is the `Galilean' one
%typical of the infinite momentum 
%frame one uses in DLCQ, and the geometrical 
%relation of this field--space Lorentz boost 
%with the `compactified' direction
%is quite suggestive of the kinematics in Matrix theory \cite{BFSS}.}
%
Since generically $Z^I_-$ transforms under \refb{e10}, the
$Z^I_-$=constant slice of the moduli space, spanned
by the coordinates $(Y^I_+, Y^I)$, is unaffected
by the identification \refb{e10}
and we get the space $R^8\times 
(R^8/\ZZZ_2)$. This is the correct moduli space for a pair
of M2-branes.\footnote{In contrast, if we had chosen to
`forget' about the $Z^I_-$ field and examined the action
on the $(Y^I_+, Y^I)$ plane, we would have 
 the identification $Y^I\sim Y^I-Y_+^I/2$. This would lead
 to the conclusion that 
spacetime gets effectively
compactified in the direction of the v.e.v. of $Y^I$ along
$Y^I_+$, with a radius 
$R\propto\sqrt{Y^I_+Y^I_+}$. This would not coincide
with the moduli space of a pair of M2-branes.}
In particular we see that the apparent $Y^I_+$ dependence
of the moduli space has disappeared so that it is now
possible to interprete $Y^I_+$ as the center of mass
coordinates of the M2-branes in eleven dimensional
flat non-compact space-time.
The moduli space does have a singularity at $Y^I_+=Y^I=0$
since this is a fixed point of the
orbifold group, but this is only to be expected since at this point
the massive fields become massless and our approximation
breaks down. The important point however is that 
this singularity does not generate a conical defect on the
$Z^I_-$=constant slice; to see this
singularity we actually have to be on the $Y^I_+=Y^I=0$ point.
We also see that the $Z^I_-$=constant condition is preserved by the
transformation \refb{e10} if we are on the subspace
$Y^I = n Y^I_+/4$. The transformation \refb{e10} takes
a point $(Y^I= n Y^I_+/4, Y^I_+)$ on this subspace to the point 
$(Y^I = -n Y^I_+/4, Y^I_+)$. Thus on the
$Z^I_-$=constant slice these two points must be
identified. But this is already guaranteed by the identification
under the exchange symmetry $Y^I\to - Y^I$; hence this does
not lead to any additional identification in the $(Y^I, Y^I_+)$
space.

%Furthermore the codimension eight subspace 
%$Y^I=n Y^I_+/4$ is identified with the codimension eight
%subspace $Y^I = (n - 2k) Y^I_+/4$ for $n,k\in\ZZZ$ via the
%transformation $Y^I\to Y^I - k Y^I_+/2$, but again the
%effect of these identifications is not visible outside these
%subspaces and could disappear in the full quantum theory.
%The full moduli space including the $Z^I_-$ coordinate
%does have a conical defect at $Y^I_+=Y^I=0$ which can be felt
%from far, {\it e.g.} by parallel transporting a vector along the
%closed curve generated by the identification 
%\refb{e10}. However this would not affect the physical 
%sector of the theory.

Our analysis can be easily extended to include the fermions.
It can also be generalized   to theories with $SU(N)$ gauge group 
instead of $SU(2)$ gauge groups. In the latter case we shall
have $(N-1)$ scalars $Y^I_k$ ($1\le k\le N-1)$
for each $I$, besides the fields
$Y^I_\pm$. For each $k$ there will be a separate global
identification of the form given in \refb{e10.1}, \refb{e10}.
However since the $Z^I_-$=constant slice is not invariant
under these transformations, this global identification
will not have any effect on the physical moduli space
(except at isolated subspaces like $Y^I_+=Y^I_k=0$.) In this case
$Y^I_+$ can be interpreted as the center of mass degrees of
freedom of $N$ M2-branes, while the coordinates
$\{Y^I_k\}$, spanning the space $(\RRR^8)^{N-1}/S_N$,
can be interpreted as the relative separation between the
branes.

Our analysis is a small step towards establishing that the
Lorentzian three algebra theories could describe the theory of
M2-branes after removal of a set of
ghost fields. In order to
establish this beyond doubt 
we need to understand how to decouple the eight ghost-like
scalars and their superpartners in the full theory and not just in the
moduli space approximation. 
Is there any way at all that we can study this question?
We suggest the following possibility: we can remain on the
coulomb branch but try to systematically study the higher
derivative correction to the low energy effective action due to
the result of integrating out
the massive modes. 
By staying in appropriate region of the moduli space where the
massive modes are sufficiently heavy,\footnote{Since the theory
is conformally invariant there is no absolute notion of
heaviness of the masses, but the relevant expansion parameter
is the ratio of the momentum carried by the moduli fields to the
masses of the heavy particles.}
it may be possible to
reorganise the perturbation expansion so that we can systematically
compute higher derivative corrections to the effective action
as an expansion in inverse powers of these masses. We can then
try to examine the effective action of the
massless fields obtained this way to see if it is possible to decouple
the ghost fields from the theory.
Furthermore after suitable choice of coordinates on the
moduli space the physical theory must be independent of the
coordinates representing the center of mass degrees of
freedom.\footnote{It is of course possible
that we must take into account the truncation of the theory before
studying quantum corrections; nevertheless examining the
effective action in the untruncated theory might provide some
clue.}
To this end we also note that effective action of massless
fields on the coulomb branch could facilitate comparison
between different proposals for the theory of multiple M2-branes.
For example recently ref.\cite{0806.1218} 
made an alternate proposal
for the theory of multiple M2-branes in which the full $\NN=8$
superconformal invariance of the theory is 
not manifest, but where the
theory is manifestly unitary. 
This theory also lacks a coupling constant, but may admit a
systematic expansion in ratios of 
momenta and vacuum expectation values of scalars.
If such computations are possible in both theories then
comparison of these effective actions could give us a
better understanding of the relationship between these different
approaches.

{\bf Acknowledgement}: A.S. would 
like to thank Shamik Banerjee, Jose Figueroa-O'Farrill,
Jaydeep Majumder and Masaki Shigemori for useful discussions.

\end{document}